\documentclass[aps,prl,twocolumn,groupedaddress]{revtex4}

\usepackage{graphicx,psfrag}
\newcommand{\be}{\begin{equation}}

\newcommand{\ee}{\end{equation}}

\newcommand{\bea}{\begin{eqnarray}}

\newcommand{\eea}{\end{eqnarray}}

\newcommand{\ba}[1]{\begin{array}{*{#1}{c}}}

\newcommand{\ea}{\end{array}}

\newcommand{\s}{s}

\newcommand{\xip}{\xi_\parallel}

\begin{document}

\title{Jamming Percolation and Glass Transitions in Lattice Models}

\author{Cristina Toninelli}

\email[]{cristina@corto.lpt.ens.fr}
\affiliation{Lab. Physique Th{\'e}orique, Ecole Normale Sup{\'e}rieure,
  24, rue Lhomond 75005 Paris FRANCE}

\author{Giulio Biroli}

\email[]{biroli@cea.fr}

\affiliation{Service de Physique Th{\'e}orique, CEA/Saclay-Orme des Merisiers,
F-91191 Gif-sur-Yvette Cedex, FRANCE }

\author{Daniel S. Fisher}

\email[]{fisher@physics.harvard.edu}

\affiliation{Lyman Laboratory of Physics, Harvard University, Cambridge, MA 02138, USA}

\begin{abstract}

A new class of lattice gas models with trivial interactions but constrained dynamics are introduced. These are proven to exhibit a dynamical glass transition: above a critical density, $\rho_c$, ergodicity is broken  due to the appearance of an infinite spanning cluster of jammed particles. 
The fraction of jammed particles is discontinuous at the
transition, 
while in the unjammed phase 
 dynamical correlation lengths and timescales diverge as
 $\exp[C(\rho_c-\rho)^{-\mu}]$.
 Dynamic correlations display two-step relaxation similar to glass-formers and jamming systems.

\end{abstract}

\maketitle

In the majority of liquids dramatic slowing down occurs upon
super-cooling.  In a rather small temperature window, typically from the
melting temperature $T_m$ to about $2 T_m/3$, the viscosity increases by $14$
orders of magnitude and the relaxation becomes complicated:
non-exponential and spatially heterogeneous
\cite{DeBenedettiStillinger}.  Similar features are observed in soft
materials, such as colloidal suspensions and more generally in non-thermal ``jamming"
systems \cite{Trappe}.  Despite a great deal of effort, these
remarkable phenomena,  associated with ``glass transitions" are
still far from understood.  Even the most basic issues remain open: is
the rapid slowing down due to proximity to a phase transition? Is
this putative glass transition static or purely dynamic \cite{Krauth}?
Experimental results make it clear, however, that {\it if} an ideal glass
transition does exist it should have some peculiar features: the density
auto-correlation function $C(t)$ 
should exhibit a lengthening plateau that, at the
transition, extends out to infinite times. Thus, the Edwards-Anderson order 
parameter, defined as the infinite time limit of $C(t)$, 
will be discontinuous at the transition. But this discontinuity should be accompanied by a critical divergence of the relaxation time. And, contrary to usual
critical slowing down, the relaxation time should diverge exponentially, as in the  Vogel-Fulcher law
\cite{DeBenedettiStillinger}.  Long-range spatial correlations, if they exist at all, must be very subtle. These unusual properties present major theoretical challenges: whether or not there is a true transition there is no ``standard" framework to start from. 
There are promising results for models on Bethe
lattices \cite{LGM} and for some with long-range
interactions \cite{Bernasconi}. But the quest for {\it models with short range 
interactions and no quenched disorder} that are {\it simple} enough to be analyzed and can
be shown to have a glass transition, namely a transition with the basic
properties discussed above,
is still open, in spite of much effort.

In this paper we
introduce  the first examples of such models \cite{foot,Liu}.
These are  {\it
kinetically constrained models} (KCMs): 
stochastic lattice gases with no static interactions, except hard core,
but constrained dynamics \cite{ReviewKLG}. The elementary moves
are particle jumps for conservative
models and birth/death moves for non conservative models. 
  Whether a move can
occur depends on the nearby configuration and is non-zero
only if some local constraints are satisfied. These kinetic
constraints can radically change the dynamical behavior and typically induce
glassy phenomenology\cite{ReviewKLG,KA,Jackle}.
For some KCMs the dynamics becomes so slow at high density or low temperature, 
that they have been conjectured to undergo a true glass transition.
The simplest examples are Kob-Andersen models on a square lattice (SKA)
\cite{KA}, where particles can move if and only if they have no more than two nearest
neighbors both before and after the move. Although the analogous model
on a Bethe lattice \cite{Bethe,KATBF} does have a jamming transition,
we have shown  previously in  \cite{KATBF} that the SKA and a broad class of generalizations on
hypercubic lattices cannot have ergodicity breaking transitions:
in any finite dimension
  the relaxation time diverges
 --- in many cases in a super-Arrhenius way --- only at the close packing
density ($\rho=1$).

\begin{figure}[bt]
\psfrag{a}[][]{a)}
\psfrag{l}[][]{$\ell$}
\psfrag{l2}[][]{}
\psfrag{l1}[][]{$\ell$}
\psfrag{c}[][]{c)}
\psfrag{b}[][]{b)}
\centerline{
 \includegraphics[width=\columnwidth]{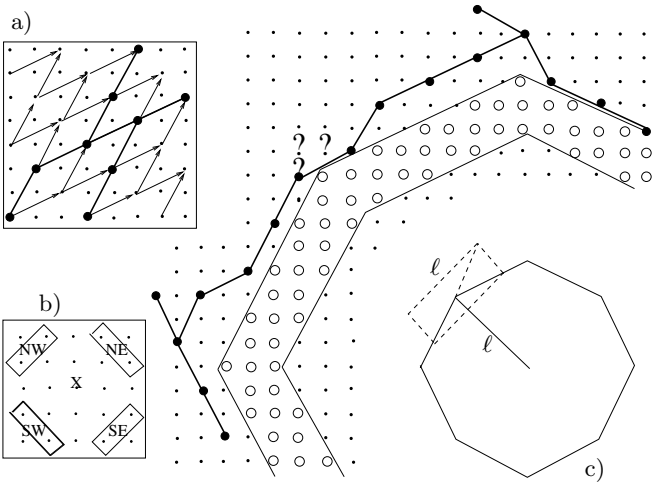}
%\includegraphics[width=\columnwidth]{9f}
%  \hspace{.3cm}
%  \includegraphics[width=0.45\columnwidth]{fignew15bis.eps}
}
\caption{a): Sites 
%(small dots) 
connected by arrows form one of the graphs on which directed percolation can form frozen clusters: e.g. the set of occupied
sites (big dots) shown. 
 b)Site X and its NE,NW,SW,SE pairs of neighbors. c) Portion of an  empty
 octagonal annulus.  The interior of the annulus, as any finite region
 surrounded by a double frame of vacancies, can be eaten away. Whether the vacant region can expand is determined by the three key
 sites indicated by question marks.  If one of these belongs to an occupied DP path in the 
%direction parallel to the adjacent side of the
% annulus 
NE direction which is anchored on two perpendicular DP paths, as shown, it blocks the expansion. 
 %here 
Inset: the full octagon. A  necessary condition for the octagon not to be
expandable of one step in the NW direction is that a DP path spans the dotted rectangle. }
\label{directed1}
\end{figure}

 But this is not the only possible behavior. We here
 introduce a new class of KCMs that do
 exhibit a jamming transition at a non-trivial critical density,
$\rho_c$ on finite dimensional lattices.\\ 
For simplicity we focus on one of the simplest: a
square lattice model {\it without} particle conservation; vacancies can loosely be thought of as ``free volume" which need only be conserved on average. At the end, we discuss generalizations to
both higher dimensional and conservative
models. 
The stochastic dynamics is as follows: 
An occupation variable at site $x$ cannot change if $x$ is  {\it
blocked} along {\it either} of the diagonal directions, as defined below. Unblocked sites change
from occupied to empty and from empty to occupied with rates $(1-\rho)$
and $\rho$, respectively.  Thus detailed
balance is satisfied with the trivial product measure with density $\rho$.
The blocking is determined by  the eight fourth-nearest-neighbor sites of $x$. Denote {\it pairs} of these the north-east (NE),
north-west (NW), south-east (SE),
and south-west (SW) pairs as in Fig. \ref{directed1} b). Site $x$ is blocked if {\it either} at least one of the NE
sites {\it and} at least one of the SW sites is occupied, {\it or}
at least one of the SE
sites {\it and} at least one of the NW sites is occupied. Blocking can thus be along either the NW-SE or the NE-SW diagonals.
As the distance to the blocking neighbors resembles a knight's move in chess, we call this the "knights model".\\
If a site cannot be unblocked even by first emptying with allowed moves an arbitrarily large number of other sites, we call the site {\it frozen}. Any finite cluster of particles cannot be frozen: one
can always unblock all sites by emptying  from the perimeter in (see Fig. \ref{directed1}c).
A crucial question is  whether an infinite spanning cluster of frozen sites 
exists in infinite systems. We will call this problem {\it jamming percolation}; it is akin to bootstrap percolation \cite{Aizenmann}.

We will show the following results (which can be 
proved \cite{TBF2long}): 
(i) with blocked or periodic boundary conditions
on $L\times L$ squares, there exist configurations with system-spanning
clusters of frozen  sites; (ii) on infinite lattices,
below a critical density, $\rho_c$, there are no infinite
frozen clusters; while (iii) above $\rho_c$, there is an infinite
cluster of frozen particles that occupies a non-zero fraction,
$\phi_{\infty}$,  of the area; (iv) $\rho_c$ coincides with the critical
density for directed  site percolation (DP) on a square lattice; (v) 
$\phi_\infty$ is discontinuous at
$\rho_c$; (vi) below $\rho_c$, there is a crossover length $\Xi(\rho)$: squares of size $L<<\Xi$ are very likely to have a frozen
cluster, while for $L>>\Xi$, the probability of a frozen cluster
falls-off exponentially; (vii) as $\rho$ increases to
$\rho_c$, $\Xi$ diverges exponentially rapidly, as
$\log \Xi\sim(\rho_c-\rho)^{-\mu}$ with $\mu\cong 0.64$ related to DP exponents; (viii) the relaxation time
diverges as $\Xi$ or faster.  
Thus even though the critical density is the same as for DP, the behavior is completely different.

We first show that both an unfrozen and a frozen phase exist.  At
sufficiently low densities, the occupied sites will not percolate via
connections up to fourth neighbors: the resulting finite clusters can always
be unblocked from their perimeters. Thus at low densities,  
 frozen clusters will not occur.
In contrast, at high densities, spanning  frozen clusters occur.
Consider site directed  percolation with directed links that 
connect a site to its two NE (fourth) neighbors as in
Fig. \ref{directed1} a).
Infinite directed paths of occupied sites exist  for $\rho\ge p_c^{DP}\cong 0.705$
(the critical threshold for conventional site DP on a square lattice \cite{reviewDP}). These clusters of sites are
frozen since  each has
at least one occupied fourth-neighbour in both NE and SW directions.
% therefore it is blocked along the NE-SW diagonal. 
Thus, $\rho_c\le p_c^{DP}$.
% is an upper bound for the critical density of our model.

For the above argument it was sufficient to use blocking along  just
one of the two diagonal directions.
But, due to blocking in the perpendicular diagonal direction, 
typical frozen configurations do {\it not} resemble DP clusters:  they consist of short DP paths that terminate at each end in a T-junction with a  DP path in
the perpendicular direction.  Thus  large regions can be frozen even if they are not spanned by DP clusters.   

\begin{figure}[bt]

\psfrag{l0}[][]{$\ell_0$}
\psfrag{a}[][]{a)}
\psfrag{s}[][]{{\small{s}}}
%\psfrag{A}[][]{{\tiny{A}}}
%\psfrag{B}[][]{{\tiny{B}}}
%\psfrag{C}[][]{{\tiny{C}}}
%\psfrag{D}[][]{{\tiny{D}}}
\psfrag{b}[][]{b)}
\includegraphics[width=.98\columnwidth]{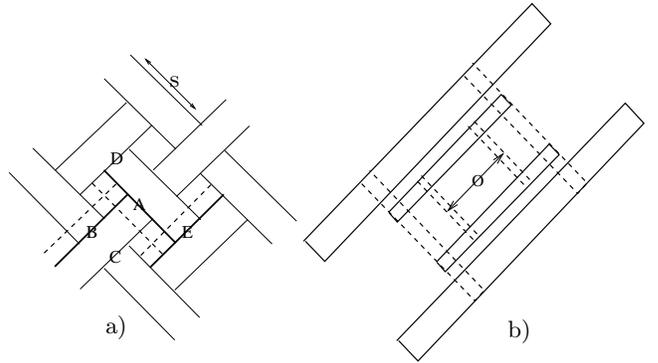}
\caption{a) Frozen structure made of DP paths of length $s$ (represented by
  straight continuous lines) intersecting at T-junctions. The  A path anchors one end of the B and C paths, while its ends are anchored by the D and E paths.  This anchoring is retained  if 
 path A is  displaced until the nearby dotted line, even if  
   the B and E  paths are similarly displaced no further than their corresponding dotted lines.  Thus the structure is frozen if all the rectangles --- shown and unshown --- formed by the solid and dashed lines are spanned lengthwise by DP paths
 b) Two sequences of intersecting
rectangles, ${\cal{R}}_i$, that when spanned length-wise by directed paths, make the central site, $O$, frozen.  For clarity, the rectangles in one direction are shown with dashed lines. }
 \label{lower}
\end{figure}

An explicit construction is instructive.
Consider a  structure built of DP paths of length, $s$,
intersecting at T-junctions
as in Fig.\ref{lower} a). This structure 
does not contain any long 
DP cluster, yet it is frozen. And, crucially, there  exists a similar frozen cluster
as long as 
each  DP path remains inside a nearby 
rectangular
region of size $s\times s/6$ (see Fig. \ref{lower} a). Therefore 
the probability of the system being frozen  is bounded from below by the
probability that {\it all} these rectangles are spanned.   
This will be substantial if the DP spanning probability of each such rectangle is very high.

 The crucial needed property  follows from the anisotropy of critical DP clusters: a cluster of length $\s$ typically  has transverse dimensions  of order  $\s^\zeta$ with $\zeta$, (often called $1/z$) the anisotropy
 exponent, $\zeta\cong 0.63$ \cite{reviewDP}.
Therefore an $\s\times b\s$ rectangle with $s$ in the parallel direction can be cut into slices of width $\s^{\zeta}$
such that for each slice the probability of DP spanning paths 
is substantial for $\rho\geq p_c^{DP}$. 
Therefore, even {\it at} $p_c^{DP}$, the  probability $r_b(s)$ of {\it not} having {\sl any} DP crossing in a large $\s\times b\s$ rectangle is $r_b(s) < {\cal O} [\exp(-2Cb  \s^{1-\zeta})]$ (with $C$ a constant).

What happens just below $p_c^{DP}$?  The above argument will still hold for
 rectangles that are are of order the DP parallel correlation length, $\xip$.
%:
% these will be spanned with probability $1-\epsilon$ with $\epsilon\sim
% r_{1/6}(\xip)$.  
This, together with the previously explained construction of a frozen structure, implies that an $L\times L$ square is likely to have a frozen structure built of DP clusters of length $\xip$ if $r_{1/6}(\xip) L^2/\xip^2\ll 1$. For $L<\Xi_< \sim
 \xip\exp(C_<\xip^{1-\zeta})$ --- a lower bound on $\Xi$ --- squares will thus
 contain frozen clusters with high probability in spite of the rarity  of  DP clusters with length larger than $\xip \ll \Xi$ .  

We now need to show that below $p_c^{DP}$, sufficiently large squares
 are {\it unlikely} to contain frozen clusters. Again, this can be done by construction --- now of unfrozen regions. Consider an infinite
 system within which is an octagonal annulus of radius (center to NW) $\ell$ 
 that is
 completely empty.  
Whether or not this empty region can be expanded depends crucially on 
 three key sites in each diagonal corner, say, the NW corner, as shown in
 Fig. \ref{directed1} c). If all three key sites are empty or emptyable
--- i.e. unfrozen ---  the empty annulus can be expanded along its two
adjacent (NNW and WNW) sides.  A necessary condition to have a key
site frozen is that it belongs to a DP cluster in the NE direction that is anchored at both ends, as in  Fig. \ref{directed1} c).   
%Indeed, blocking in the perpendicular (NW) direction cannot occur since the
% corresponding cluster will always  intersect the empty octagon and therefore
% it can be unblocked starting from there.  
Furthermore, in order for this anchorage to occur, it is necessary that the NE path
spans the rectangular dotted region of size
$\ell\times b\ell$ (with $b$ a constant) in Fig 1 c).
%For the annulus to be (iteratively) expanded by a factor of $\frac{3}{2}$, a necessary condition is that the $\ell\times 2\ell/3$ rectangle shown in Fig \ref{directed1} c) is not spanned lengthwise by any DP clusters
%Note that such a DP cluster is not,  by construction, ``supported" by any
% perpendicular paths on its interior side. 
For $\rho<p_c^{DP}$ DP clusters with length $\ell$ much larger than the DP
correlation length, $\xi_\parallel$, are exponentially rare and the rectangle
spanning probability  is $\sim \exp(-\ell/\xi_\parallel)$
\cite{reviewDP}.  Thus, if $\rho<p_c^{DP}$,  the probability that the annulus can be expanded out to
infinity 
by successive expansions
%factors of $\frac{3}{2}$ 
is high for $\ell >>\xi_{\parallel}$  
%(see e.g. 
\cite{ReviewKLG,Aizenmann,KATBF}.
%in the context of bootstrap percolation or 
 %for KA model).
Since the infinite system will contain a non-zero density of these empty
regions which can be expanded to unblock the whole system, we conclude that $\rho_c=p_c^{DP}$.

Estimating how rare the unblocking regions are near $\rho_c$, yelds an upper bound on
 the crossover length $\Xi$.  Starting with a small empty nucleus  the --- small --- probability, $\delta$,  that it can be expanded
 out to size $\sim\xip$  (and hence readily to infinity)  is the product of many small terms. This is dominated by $\ell \sim\xip$: from $r(\xip)$, we  obtain $\delta > 
 \exp(-2C_>\xip^{1-\zeta})$. 
 %One can reach the
  %empty octagon of size $\xip$ iteratively doubling the size of the initial empty
%nucleus with the expansion strategy outlined in the text.
%A sufficient condition to double an octagon of side $\ell$ is
%not having a crossing DP path in regions of linear size $\ell$.
%The probability of this event is of the order of $\exp(-Cl^{1-\zeta })$ with $C$ a suitable constant (as long as $\ell<\xip$).
%The product of these terms from $l$ of the order of one to
%$l\simeq \xi_{\parallel }$ is dominated by the last one. Hence,
%$\delta > \exp(-2C_>\xip^{1-\zeta})$. 
Since in an $L\times L$ square
there are $(L/\xip)^2$ roughly independent positions for such empty nuclei, some are likely to occur if $\delta L^2/\xip^2$ is not small. Thus we find that
$\Xi\le \Xi_>\sim \xip\exp(C_>\xip^{1-\zeta})$. \\
We have found upper and lower bounds for the crossover length, $\Xi$, of similar form, hence
$$
\log\Xi \sim \xip^{1-\zeta} \sim (\rho^c-\rho)^{-\mu} ~~ {\mbox {with}} ~~\mu=(1-\zeta)\nu_\parallel\cong 0.64
$$
with $\nu_\parallel \cong 1.73$ the correlation length exponent for DP \cite{reviewDP}.

We now discuss 
another peculiarity of this transition: the nature of the frozen clusters implies that the 
density, $\phi_\infty$, of  the infinite frozen cluster jumps at
$\rho_c$.
To analyze the probability that a site is frozen, 
consider an occupied site, e.g. the origin, which belongs to a
DP cluster that extends to a (small) distance  $\ell_0/2$ in both the NE and SW directions: this occurs with some probability $p_0$. Now focus on 
two infinite sequences of rectangles ${\cal{R}}_i$
of increasing size $\ell_i\times \ell_i/12$ with $\ell_1=\ell_0$,
$\ell_{i}=2\ell_{i-2}$ and intersecting as in Fig. \ref{lower}.
If each of these rectangles is spanned
lengthwise by a DP cluster, 
the origin is frozen. The probabilities of perpendicular DP paths are
positively correlated. Thus $\phi_\infty\ge p_0 \prod_i
[1-r_{1/12}(\ell_i)]^2$. 
 %Therefore, since the $\rm{Prob}[\rm{all \ spanned}\ge 1-\sum_i\rm{Prob}[\cal{R}_i\  \rm{not\  spanned}]$,  $\phi_{\infty}\ge p_0-2\sum_{i=1}^{\infty} r_{1/12}(\ell_i)$, 
%with $r_b(\ell)$   the (small) probability that an $\ell\times b\ell$ rectangle
%is {\it not} spanned. 
Since $\xip$ is infinite at $\rho_c$, as shown above, $r_b(\ell)$ decays exponentially to zero as $\exp[-Cb\ell^{1-z}]$.
Therefore, the infinite product is non-zero, giving
a strictly positive bound on $\phi_{\infty}(\rho_c)$ for any
% as soon as 
$\ell_0$. 
%is choosen sufficiently large. 
Thus, in contrast to DP or conventional percolation,
the infinite frozen cluster of jamming percolation is ``compact" --- i.e. with dimension $d$ ---
at the transition.

To supplement our predictions, we have studied $L\times L$ systems numerically. The distribution of the densities, $\phi_L$, of the frozen clusters shows two peaks with weak size dependence as found at conventional first order transitions. This is consistent with the 
predicted discontinuous behavior.
The probability that there {\it exists} a frozen cluster is substantial for
$\rho$ fifteen percent below $\rho_c$ even in our largest systems,
($L=1600$): it is thus hard to study the asymptotic critical behavior (see
\cite{Dawson} for an analogous problem in the context of bootstrap
percolation).   But in a slightly different model one can get closer to the
transition \cite{TBF2long}: these data are consistent with the  predicted $\ln\Xi \sim (\rho_c-\rho)^{-\mu}$ with $\mu\cong 0.64$, but the small range of $\ln L$ available makes the uncertainties in $\mu$ large.  

Our results on  {\sl jamming percolation} have
important implications for the equilibrium dynamics of the knights model. 
 For $\rho<\rho_c$ the fact that infinite frozen clusters do not exist
imply that the system is ergodic [proof can be done as in (\cite{KATBF} 2.5)] but dynamical correlations, such as the  density
auto-correlation function, $C(t)$, display two step temporal relaxation with a
long plateau followed eventually by a relaxation (numerical results
will be reported elsewhere \cite{TBF2long}). The  relaxation time diverges exponentially near $\rho_c$, at least as $\Xi ^{z}$, with $z\geq 1$:  This is reminiscent of the Vogel-Fulcher law found experimentally near glass transitions.
Above $\rho_c$, the plateau stretches to infinite times with the
Edwards-Anderson order parameter, $q\equiv\lim_{t\rightarrow \infty}C(t)$,
discontinuous at the transition. This follows from our results, since $q$
 is related to the density of
frozen sites.

We have seen that, even though the critical densities are the same, 
the  properties of   jamming percolation are strikingly
different from the power law behavior of directed percolation. Most of the physics is controlled by relatively short DP clusters joined together at T-junctions. The only role of long DP clusters is to prevent  very rare large unfrozen regions from unblocking their surroundings.   There is substantial universality in the primary features of the jamming percolation. This extends even to the SKA model which has frozen configurations composed of double-width
 occupied bars that terminate in T-junctions with similar perpendicular
 bars, but there is no real transition because very long bars are
 unlikely. Yet the SKA's behavior as $\rho\nearrow 1$ is similar to the
 knights model as $\rho\nearrow \rho_c$ with $1/(1-\rho)$ roughly replacing
 powers of $\xip$.
  Surprisingly,  if  the square lattice of the SKA is
 replaced by a particular complicated four-fold coordinated lattice, $\rho_c<1$ and the behavior is  similar  to the knights model.  Thus the local structure matters a lot as in real glasses.
 Note that cooperative models with a transition, in contrast to those without, 
 display two-step relaxation from the dynamics within blocked regions: this is like  beta-relaxation in glasses \cite{DeBenedettiStillinger}. 
% not unreasonable for glass transitions. 

Models with {\it particle-conserving} dynamics behave surprisingly
similarly to those without: the nature of the transition (and in some
cases the critical density)  remain the same because the slowing of
the  dynamics is dominated by the large clusters of the underlying
jamming percolation. Diffusive transport rides on top of this \cite{TBF2long,Huse,Chandler}.
In three dimensions, two natural generalizations of our jamming percolation exist: one composed of DP clusters --- which should slow down as a double exponential of $(\rho-\rho_c)^{{-\overline{\mu}}}$ ---  and the other of directed sheet-like structures which will have exponential slowing down like we have found in 2D.  The key ingredients are  kinetic constraints that enable huge jammed clusters to form out of small objects without these becoming much more common or much larger. 

For the future, the connection between our results and the jamming transition found for continuum particle systems \cite{Trappe} needs exploring.
With the  hope of increased understanding of the rapid liquid to glass
crossover observed experimentally, one should also analyze the effects of
constraint-violating processes occurring with a very low rate.  For both
glasses and granular materials, studying the non-equilibrium effects caused by
a quench or by driving forces \cite{TBF2long} is merited even in the simplest models that exhibit a jamming transition.\\
After the completion of this work a new version of a preprint \cite{Liu}
appeared in which other models with a jamming transition are introduced
and studied numerically.

\begin{acknowledgments}
We thank J. M. Schwarz, A. J. Liu for
discussions.  The numerical simulations have been performed on the
parallel computer cluster of CEA under grant p576.  GB is partially
supported by EU contract HPRN-CT-2002-00307 (DYGLAGEMEM).CT 
by EU contract HPRN-CT-2002-00319(STIPCO) and DSF by the NSF via DMR-0229243.
\end{acknowledgments}


\begin{thebibliography}{99}
\bibitem{foot} The ``North-East" KCM has an ergodicity breaking transition
\cite{ReviewKLG}, but the  constraints are non-reciprocal --- thus unphysical for glasses --- and the transition  is continuous with power-law critical behaviour.
\end{thebibliography}
\end{document}